\documentclass[twocolumn,showpacs,preprintnumbers,amsmath,amssymb]{revtex4}
\usepackage{graphics}

\begin{document}

\title{Resistive and rectifying effects of pulling 
gold atoms at thiol-gold nano-contacts} 

\author{ Ronaldo J. C. Batista$^{1,2}$ }
\email{ronaldo@fisica.ufmg.br}
\author{ Pablo Ordej\'on$^3$}
\author{Helio Chacham$^1$}
\author{Emilio Artacho$^2$}

\affiliation{$^1$Departamento de F\'{\i}sica, ICEX, Universidade 
Federal de Minas Gerais, CP 702, 30123-970,Belo Horizonte, MG, Brazil}
\affiliation{$^2$Departament of Earth Sciences, University of Cambrigde, 
Downing Street, Cambridge CB2 3EQ, UK}
\affiliation{$^3$ Intitut de Ci\`encia de Materials de Barcelona, CSIC, 
Campus de la UAB, 08193 Bellaterra, Spain}
\date{\today}

\begin{abstract}
We investigate, by means of first-principles calculations, structural 
and transport properties of junctions made of symmetric dithiolated 
molecules placed between Au electrodes. As the electrodes are pulled apart, we
find that it becomes energetically favorable that Au atoms migrate
to positions between the electrode surface and thiol terminations, with
junction structures alternating between symmetric and asymmetric. As a
result, the calculated $\emph{IV}$ curves alternate between rectifying 
and non-rectifying behaviors as the electrodes are 
pulled apart, which is consistent with recent experimental results.
\end{abstract}

\maketitle

  Stimulated by recent experiments \cite{nature,nature2,materialstoday}, 
the use of organic molecules as basic components of nanostructured circuits 
has been envisaged as a way to the further minituarization of electronic 
devices. 
  In order to achieve real devices it is required a good understanding of the 
features affecting the conductance of a single molecule attached to electrodes.
  The atomic structure of the contact region may strongly affect the 
conductance and the current-voltage ($\emph{IV}$) characteristics of a 
molecular junction. 
  For instance, the choice of different functional groups connecting the 
molecular core to the electrodes has been shown to lead to very asymmetric 
$\emph{IV}$ curves \cite{materialstoday2}.
  Surprisingly, rectification can also occur in symmetric molecules:        
in some experiments the careful manipulation of the electrode distances 
in a symmetric molecule-gold junction can result in asymmetric $\emph{IV}$
characteristics \cite{prl2002,prb99}. 
  Since the organic molecule is symmetric, the probable reason for the 
asymmetric $\emph{IV}$ curves 
is the presence of asymmetric contacts.
  Experimentally, it has been found that the force required to break a 
molecular junction of a single octanedithiol molecule attached to gold 
electrodes is similar to the value of the force required to break a Au-Au 
bond in an atomic gold chain (1.5 - 1.8~nN) \cite{rubio}. 
  This fact strongly suggests that the dithiol molecule can pull Au atoms 
off the electrodes when a external force is applied.
  Molecular dynamics simulations also indicate that a thiolate molecule can 
pull gold atoms off a stepped surface \cite{parrinello}. 
  Therefore, asymmetric contacts could be formed by stretching the molecular 
junctions which in turn could be the source of such asymmetries in the 
$\emph{IV}$ curves. 

  In this paper we have investigated the effect of the strain on the atomic 
structure and on the 
electron
transport properties of molecular junctions composed 
by symmetric dithiolated molecules attached to Au(111) electrodes. 
We find that the conductance is not very sensitive to small applied 
forces, in agreement with experimental observations made by Reichert 
{\it et al.} \cite{prl2002}.
  However, as the electrodes are pulled apart, we
find that it becomes energetically favorable that Au atoms migrate
to positions between the electrode surface and thiol terminations, with
junction structures alternating between symmetric and asymmetric. As a
result, the calculated $\emph{IV}$ curves alternate between rectifying
and non-rectifying behaviors as the electrodes are
pulled apart, which is consistent with the results of Reichert
{\it et al.} \cite{prl2002}.
\begin{figure}
\centering
\resizebox{0.4\textwidth}{!}{\includegraphics{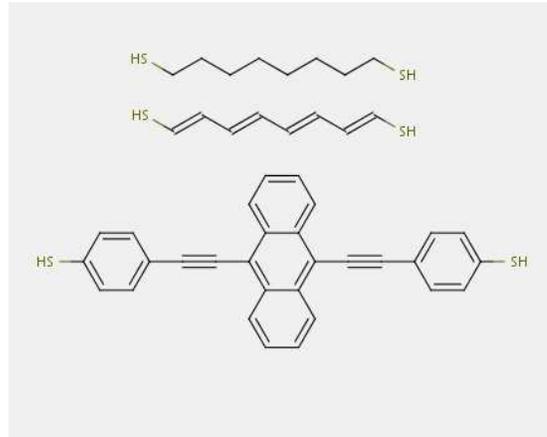}}
\caption{Alkanedithiol (Octanedithiol), Alkenedithiol (Octenedithiol),
and [9,10-Bis((2'-\emph{para-mercaptophenyl}
-ethinyl)-anthracene] (BPMEA)}
\label{fig1}
\end{figure}

\begin{figure}
\centering
\resizebox{0.4\textwidth}{!}{\includegraphics{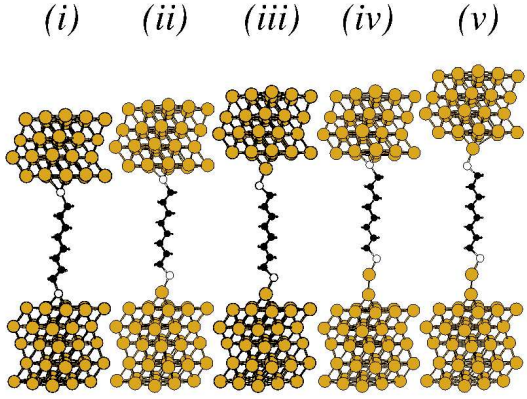}}
\caption{Studied gold-octanedithiol-gold junctions. Similar junctions 
of alkenedithiol were also studied.}
\label{fig2}
\end{figure}

  Our methodology is based on density functional theory (DFT) \cite{dft} 
within the generalized gradient approximation (GGA) \cite{pbe} and 
norm-conserving Troullier-Martins pseudopotentials \cite{pseu} as implemented 
in the {\sc Siesta} method \cite{siesta1,siesta2}. 
  The solutions of the Kohn-Sham equations are expanded as a linear combination 
of atomic pseudo-wavefunctions of finite range. 
  We used a basis set previously used to study methylthiol functionalized gold 
nanoclusters\cite{eu,eu2,garzon1}. 
  The conductance and the current are calculated within a nonequilibrium 
Green's-function formalism at the same DFT level using the {\sc TranSiesta} 
method \cite{transiesta}.

  We have considered the three organic molecules shown in Fig.~1:
an insulating alkane chain (octanedithiol); a conducting alkene 
chain (octenedithiol); and the symmetric conjugated polymer
[9,10-Bis((2'-\emph{para-mercaptophenyl}-ethinyl)-anthracene](BPMEA). 
  The molecules are connected to the gold electrodes through the sulphur 
atoms of the thiol groups, as shown in Fig.~2. 
  The electrodes are described as in Ref.~\cite{transiesta}, with
a (111) gold slab that connects to the bulk of gold leads through
their self-energy matrix.
  The thickness of the slab is 9 layers, out of which one is defined
as buffer \cite{transiesta}, and the lateral supercell repetitions used are 
3x3 and 5x5.
  We find that the 3x3 cells are large enough for the convergence of the 
total energy calculations, while the conductance calculations demand
5x5 cells for a convergence of the positions of the transmission peaks within 
0.25~eV\footnote{Artificial broadening effects due to the contact to the leads
are completely negligible, since the imaginary part of the energy used 
in the transmission
function calculation (7.34x$10^{-7}$~eV) is very small in comparison to the
width of the transmission peaks} .

\begin{figure}
\centering
\resizebox{0.45\textwidth}{!}{\includegraphics{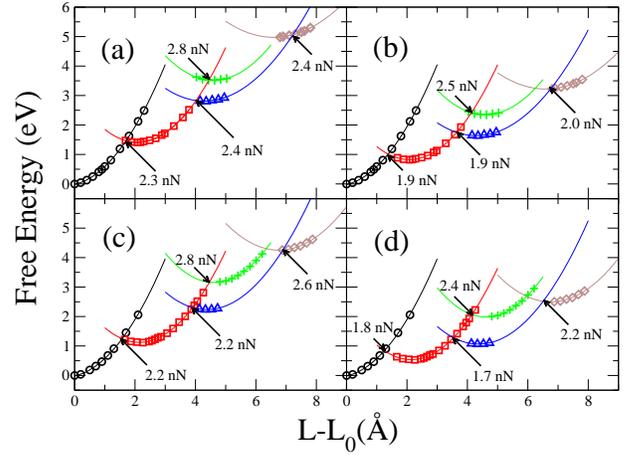}}
\caption{(Color online). Helmholtz free energies $\Omega$
of gold/molecule/gold systems as a function of the electrode-electrode 
distance. The top panels [(a),(b)] correspond to the gold/octanedithiol/gold
systems, and the bottom panels [(c),(d)] to the gold/alkene/gold systems.
In the left [(a),(c)] panels $\Omega$ was calculated with $\mu_{Au}=\mu_{bulk}$ 
and the right panels [(b),(d)] with  $\mu_{Au}=\mu_{adatom}$. 
In each panel, the curves from left to right 
($circles\rightarrow squares\rightarrow triangles\rightarrow cross\rightarrow 
diamonds$) correspond to the contact geometries from the left to the right in 
Fig.~2 ($\emph{i}\rightarrow \emph{ii}\rightarrow \emph{iii}\rightarrow 
\emph{iv}\rightarrow \emph{v}$).}
\label{fig3}
\end{figure}

  In order to address the question of the stability of additional gold
atoms between the sulphur atom and the gold surface, we considered 
the five systems shown in panels $\emph{(i)}$ through $\emph{(v)}$ 
in Fig.~\ref{fig2}: 
  In $\emph{(i)}$ the molecule (alkane in the figure) is directly connected 
to the gold slabs; 
  in $\emph{(ii)}$ an extra gold atom is placed in between one sulphur atom 
and the gold surface; 
  in $\emph{(iii)}$ extra gold atoms are found at both contacts;
  in $\emph{(iv)}$ two atoms are placed between the sulphur
and the surface in one of the contacts; 
  in $\emph{(v)}$ one contact has two extra atoms and the other contact has one.
  Similar configurations have been also considered for the alkene chain.
  The Helmholtz free energy $\Omega$ is calculated (at $T=0$) as a function
of length and compared in Fig.~3 for the systems described.
  $\Omega$ is the relevant energy for the experimental realizations of
these devices since, even if the force is measured, it is the length what 
is defined by the experiment \cite{xu}.
  In our simulations, we increase the electrode-electrode distance
in quasi-static steps by increasing the size of the unit cell in the
\emph{z}-axis. 
  In each step the lattice vectors are kept fixed and the geometry of the 
system is allowed to relax. 
  The Helmholtz free energy is calculated according to
\begin{equation}
\Omega = E_{total}-\mu_{Au}N_{Au},
\end{equation}   
where $E_{total}$ is the calculated total energy, $N_{Au}$ is the number of 
extra gold atoms, and $\mu_{Au}$ is the chemical potential of gold. 
  As a lower bound for $\mu_{Au}$ we consider 
$\mu_{Au}=\mu_{Au}^{\emph{bulk}}$, the total energy per atom for bulk gold. 
  As a plausible upper bound we choose $\mu_{Au}=\mu_{Au}^{\emph{adatom}}$,
the energy of an adatom on the Au(111) surface, i.e., the energy difference
between the Au(111) slab with and without an adatom.

  Fig.~3 displays the Helmholtz free energy as a function of the length. We also
indicates the value of the force at the intersections (the force is $F_z=- \delta \Omega / 
\delta z$ as coming from the low $L$ side).
  These force values can be seen as estimates of the forces required 
to extract atoms from the gold surface, at least in the adiabatic 
low-temperature limit.
  Indeed, the force required to extract the first atom (1.8-2.2~nN)
is in good agreement to the value obtained by Kr\"uger {\it et al.}\cite{parrinello} using Car-Parrinello molecular dynamics simulations. These extracting forces are not very different for alkane and alkene 
junctions. 
This suggests that the forces required to pull off Au atoms from the surface
do not depend much
on the organic molecular core, and one could expect similar 
values for other dithiolated molecules.
  The structures $\emph{(iii)}$ and $\emph{(iv)}$ in Fig.~\ref{fig2}
have the same number of atoms. 
  Their Helmholtz free energies shown in Fig.~\ref{fig3} indicate that 
structure $\emph{(iii)}$ is the most stable one, by 0.7~eV. 
  Therefore, considering energetics aspects only, the results of Fig.~3 
would indicate the following sequence of junctions during the pulling of 
the electrodes: $\emph{(i)}$, symmetric; $\emph{(ii)}$, asymmetric;
$\emph{(iii)}$, symmetric; $\emph{(v)}$, asymmetric. 
  Since an additional gold atom between the sulphur and the electrodes 
generates asymmetries in the \emph{IV} characteristics of dithiol junctions, 
as we shall show below, the sequence of symmetric and asymmetric contacts 
described above is consistent with the observation of both symmetric and 
asymmetric \emph{IV} curves by Reichert {\it et al.} \cite{prl2002} for a 
BPMEA attached to gold slabs.
  In that experiment, starting with a symmetric \emph{IV} they increased
the inter-electrode distance obtaining an asymmetric \emph{IV}. 
  After further manipulations they recovered the symmetric \emph{IV}
followed again by asymmetric curves.

\begin{figure}
\centering
\resizebox{0.4\textwidth}{!}{\includegraphics{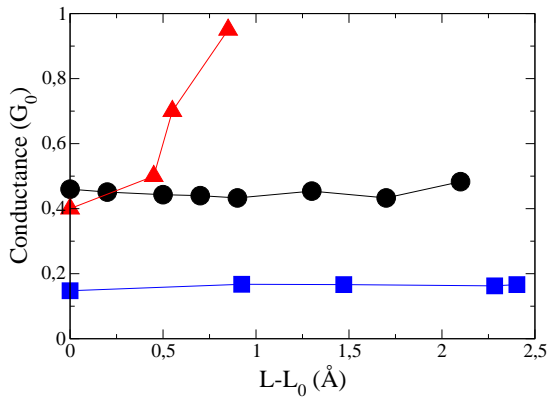}}
\caption{(Color online). Conductance through octenedithiol 
as a function of the inter-electrode separation:
black circles correspond to the Fig.~2$(i)$ junction, in which the molecule 
is directly connected to the Au(111) surfaces; blue squares
are for the Fig.~2$(ii)$ junction, in which an extra gold atom is placed
between a sulphur atom and the electrode; red triangles are for the 
Fig.~2$(i)$ junction, for unrelated electrodes.
}
\label{fig4}
\end{figure}

  Fig.~\ref{fig4} shows the conductance of the octenedithiol as a 
function of the distance between electrodes for the first and second 
junctions in Fig.\ref{fig2}, i.e., with and without an extra gold atom
(circles and squares, respectively).
  There is no appreciable change in conductance for any of the systems
over the quite substantial extensions displayed, which is consistent with 
what is observed experimentally by Reichert {\it et al.} \cite{prl2002}. 
  They find essentially unaltered \emph{IV} characteristics upon extension 
of up to a few \AA, and suggest that the strain is relieved by the soft gold 
bonds. 
  To test this hypothesis we performed calculations keeping the geometry 
of the gold electrodes fixed, relaxing only the molecule for each elongation. 
  In this non-relaxed case, the conductance of the alkene chain increases from 
approximately 0.4~$G_{0}$ to almost 1~$G_{0}$ when the electrode distance is 
increased by only 0.6~\AA, as shown Fig.~\ref{fig4}. This shows that the
weak dependence of the conductance with elongation is due to the 
relaxation of the electrode geometry near the contact region.
\begin{figure}
\centering
\resizebox{0.5\textwidth}{!}{\includegraphics{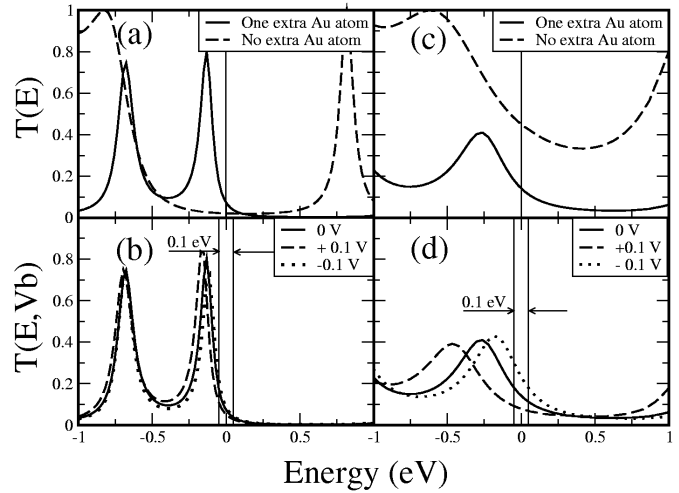}}
\caption{(Color online). Total transmission coefficient versus energy.
(a) and (b) are for BPMEA, (c) and (d) are for octenedithiol.
The upper panels compare the junctions without (dashed lines) and
with (continuous lines) an extra gold atom at a thiol-electrode junction.
The lower panels show the effect of the applied bias on the extra-atom
junctions in both cases. In lower panels, the energy origin corresponds to the average of the chemical potentials
in the left and right electrodes, $\frac{\mu_{L}+\mu_{R}}{2}$.}
\label{fig5}
\end{figure}

\begin{figure}
\centering
\resizebox{0.35\textwidth}{!}{\includegraphics{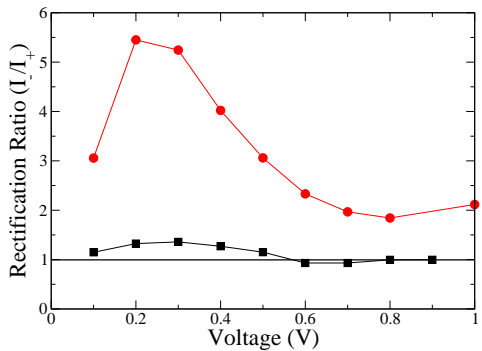}}
\caption{(Color online). Rectification ratio $\frac{-I}{+I}$ for the
octenedithiol junctions in Fig.~2$(i)$ (direct contact to the electrodes,
black squares) and Fig.~2$(ii)$ (one extra Au atom in one of the
thiol junctions, red circles).}
\label{fig6}
\end{figure}

  In contrast with the small effect of elongation on the conductance
  described above, the effect of extra Au atoms in the conduction
  of the molecular junction is very large, as indicated in Fig. \ref{fig4}.
  The figure shows that the conductance of the Au/octenedithiol/Au
  junction is reduced by a factor of three upon the the inclusion
  of an extra Au atom between one of the molecular thiol termination
  and the electrode.
  This effect is further explored in Fig.~5 for BPMEA [(a) and (b)] and for 
octenedithiol [(c) and (d)].
  The upper panels show the effect of the additional gold atom in the low
bias conductance, while the lower panels show the asymmetric bias dependence
for the junctions with the extra atom.
  The effect of the extra atom on the conductance is significant, 
decreasing G to approximately one third of its original value for the alkene, 
while it increases from 0.02~$G_{0}$ to 0.07~$G_{0}$ in the case of BPMEA. 
It is worth to note that the calculated values of conductance of the 
BPMEA junction without an additional gold atom($0.02 G_{0}$) is in 
qualitative agreement with the experimental result obtained by Reichert et al\cite{prl2002},
$\sim 0.01 G_0 $.
  The applied bias produces a shift of the spectral features 
in the transmission, which gives rise to asymmetry in the \emph{IV}
characteristics.
  Under applied 
bias 
values of 0.1~V and -0.1~V, we obtain currents 
of 1.2~$\mu$A and 0.7~$\mu$A, respectively, for BPMEA.
  Fig.~\ref{fig6} shows the rectification ratio, $\frac{-I}{+I}$,
for the alkene junction with and without the extra atom, displaying
a clear rectification in the former case (circles).
  The much smaller rectification observed in Fig. 6 for 
the case without extra Au atoms is a result of the absence of 
a left-right symmetry in the device (the molecule has inversion symmetry while
the contacts have reflection symmetry), which gives rise to a
small intrinsic effect.
  The introduction of the extra atom produces an increase of
an order of magnitude in the rectification.

  In conclusion, we find that the forces normally present in 
experiments with molecular nanojunctions do not directly affect
the conductance characteristics through these devices due to the
softness of the gold leads. 
  They can, however, stabilize extra gold atoms at the electrode-thiol
junctions, which do change the conductance and give rise to asymmetry and
appreciable rectification effects, which is in qualitative agreement
with recent experimental results. 
Our results concerning the mechanical properties of the junctions 
are weakly dependent on the actual molecule between the electrodes,
being mainly determined by the thiol-gold chemistry.

\begin{acknowledgments}
We acknowledge support from the Brazilian agencies CNPq, FAPEMIG and CAPES.
PO acknowledges funding from spanish MEC (BFM2003-03372-C03-01) and
Generalitat de Catalunya (SGR-2005 683).
\end{acknowledgments}


\end{document}